\documentclass[11pt]{article}
\usepackage{amsmath}
\usepackage{amssymb}
\usepackage{graphicx}

\begin{document}

\setcounter{figure}{0}
\setcounter{table}{0}
\setcounter{footnote}{0}
\setcounter{equation}{0}

\noindent {\Large\bf Outstanding Pulkovo latitude observers Lidia Kostina\\ and Natalia Persiyaninova\footnote{Presented
at the Journ\'ees 2019 ``Astrometry, Earth rotation and Reference systems in the Gaia era'', Paris, France,
7-9 Oct 2019, https://syrte.obspm.fr/astro/journees2019/}}

\vspace*{0.5cm}
\noindent {\large\bf Zinovy Malkin, Natalia Miller, Tatiana Soboleva}\\
\noindent {Pulkovo Observatory, St. Petersburg, Russia, e-mail: malkin@gaoran.ru}\\

\vspace*{0.5cm}
\noindent {\large\bf ABSTRACT.}
Lidia Dmitrievna Kostina and Natalia Romanovna Persiyaninova left a bright mark in the history of the Pulkovo
Observatory, as well as in the history of the domestic and international latitude services.
In the first place, they were absolute leaders in the latitude observations with the famous zenith telescope ZTF-135.
In 1954-2001, they obtained together more than 66'000 highly accurate latitudes, which make about 2/3 of all
the observations made by 23 observers with the ZTF-135 after the WW2.
They also provided a large contribution to investigation of the instrumental errors, methods of the data analysis,
developing of the observing programs. Their results in studies of the latitude variations and polar motion were also
highly recognized by the community.

\vspace*{1cm}

Pulkovo astrometrists Lidia Dmitrievna Kostina (Dec 8, 1926 -- Jun 4, 2010) and Natalia Romanovna Persiyaninova 
(Aug 26, 1929 -- Jan 16, 2003) left a bright mark in the history of the Pulkovo Observatory, as well as 
in the Russian and international latitude services. 

First, Lidia Kostina and Natalia Persiyaninova have been known as outstanding observers at the famous zenith telescope 
ZTF-135 (originally called Large Pulkovo Zenith Telescope with the lens diameter of 135~mm constructed in 1899--1904
in the mechanical workshop of the Pulkovo Observatory by Heinrich Andreevich Freiberg).
It is known that ZTF-135 was recognized as the best zenith telescope in the world. 
However, the glory of this instrument was deserved not only by the art of its creator, but also by the highly 
qualified work of the observers. 
The Pleiades of the remarkable observers of the ZTF-135, the brightest representatives of which 
were Natalia Persiyaninova and Lidia Kostina.
Together they obtained 39\% of all observations on this instrument in its more than a century history, including 59\% 
of all post-WW2 observations. They were the absolute leaders among Pulkovo (and most likely all domestic) observers
in the number of observing nights, as well as ZTF-135 was the leader among Pulkovo instruments. 
They were repeatedly awarded with the first and other prizes in the ``Best Pulkovo observer of the year'' competition.

Lidia Kostina and Natalia Persiyaninova also made a very large contribution to the research of the instrument,
to the theory and practice of observation processing, interpreting changes in latitude and studying the motion 
f the Earth's pole, determining astronomical constants. In these fields, they deserved great prestige among their
colleagues too.

\begin{figure}
  \begin{minipage}[b]{0.4455\textwidth}
    \includegraphics[clip,width=\textwidth]{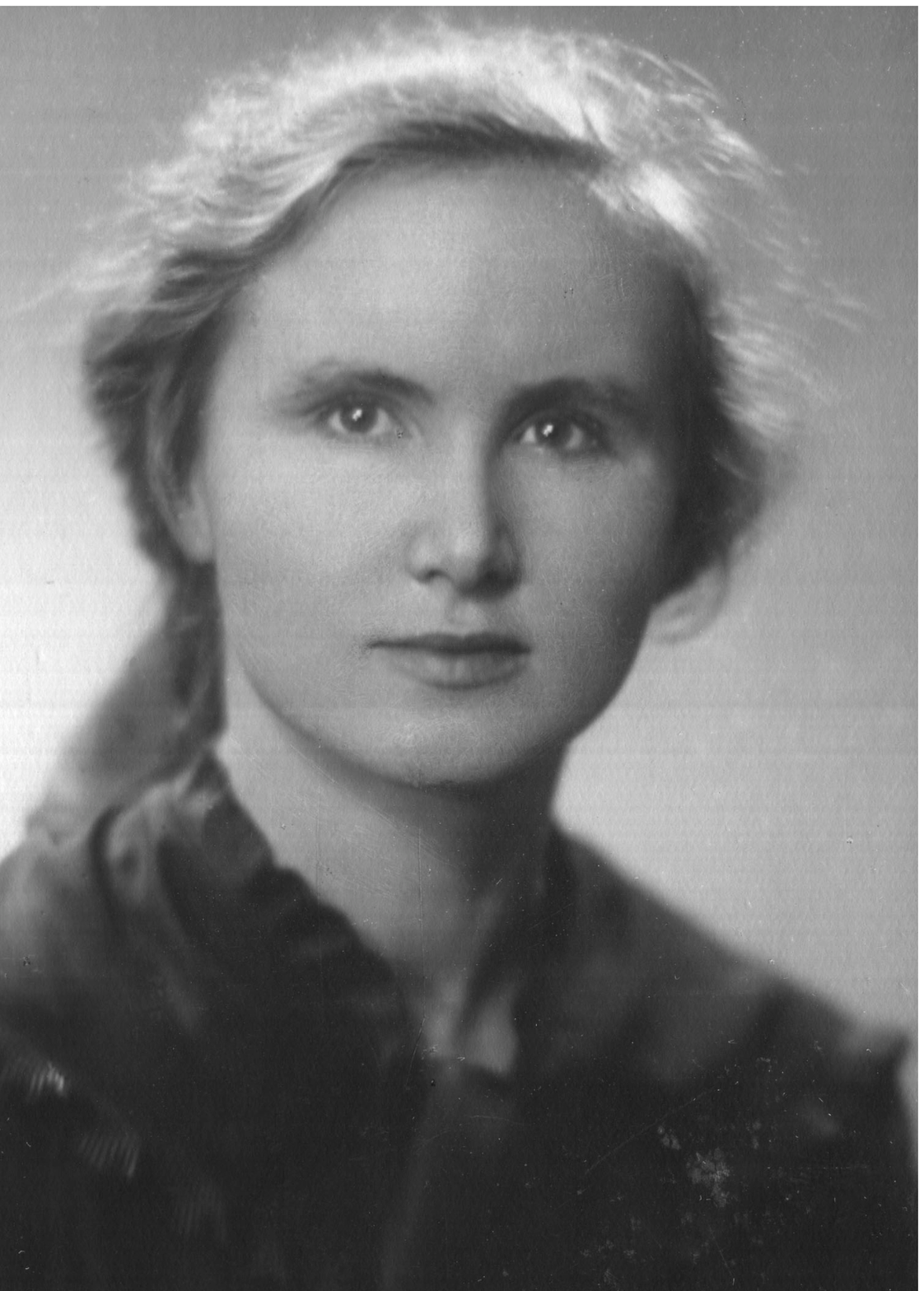}
    \caption{L.D.Kostina, 1956.}
  \end{minipage}
  \hfill
  \begin{minipage}[b]{0.5345\textwidth}
    \includegraphics[clip,width=\textwidth]{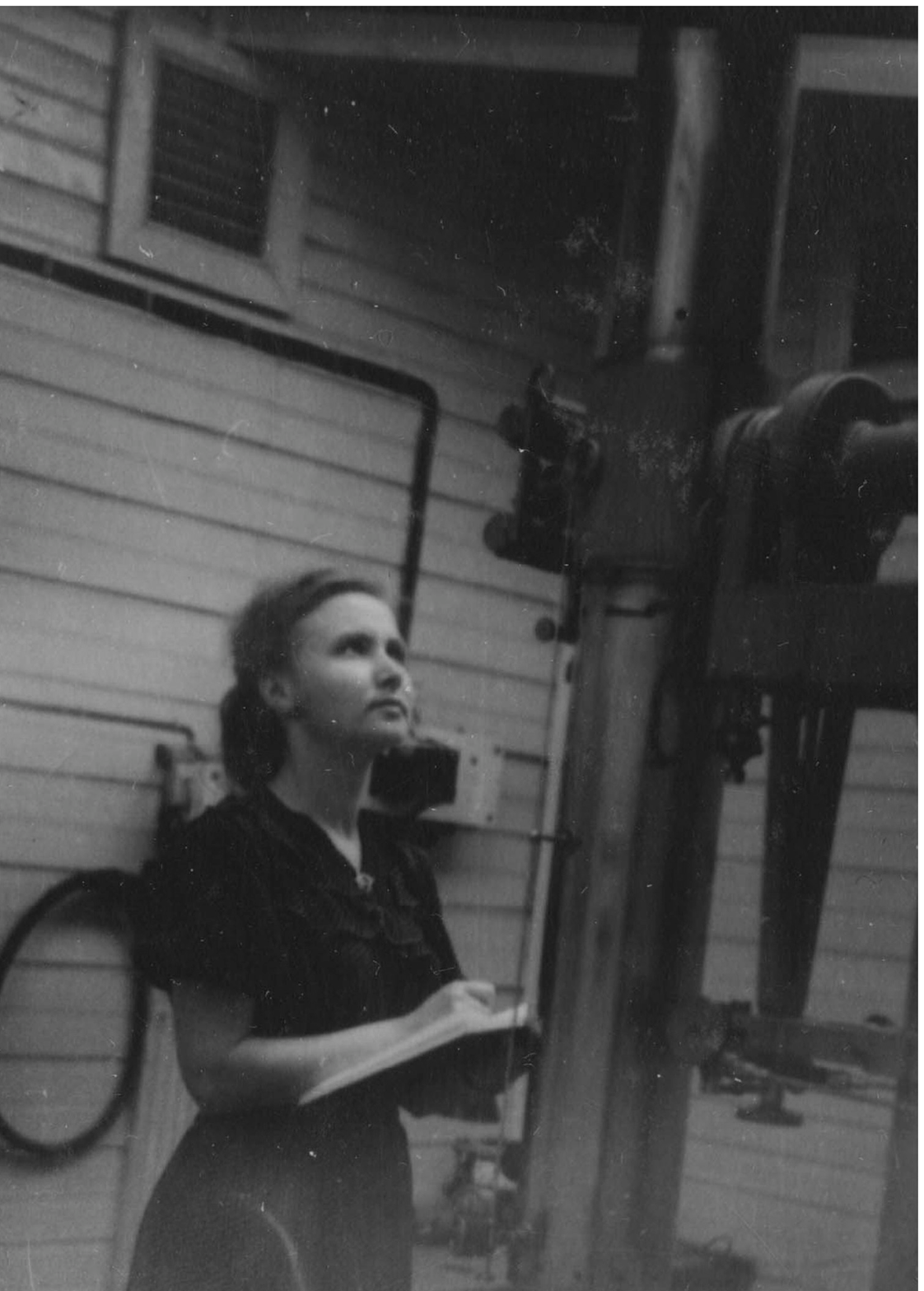}
    \caption{L.D.Kostina at the ZTF-135, 1956.}
  \end{minipage}
\end{figure}

In the year of graduation of Lidia Kostina from the Leningrad University (1950), the Pulkovo Observatory was still 
being restored. 
She joined the Northwest Airborne Surveying Enterprize, but soon became a graduate student at the Pulkovo Observatory. 
After completing the course in 1955, she became an employee of the Department of Astronomical Constants and the Earth's 
Pole Movement, ZTF-135 group, where she worked for the rest of her scientific life. 
In 1959, she defended a dissertation on the topic ``Determination of Right Ascension of FK3 Near-Polar Stars''.

Although the restored Observatory was officially opened in May of 1954, regular observations with ZTF-135 started 
in September 1948. 
Lidia Kostina joined the latitude observers group and immediately became involved in  processing of the  observations. 
Since April 1956, she was also included in the observing schedule. 
It is worth recalling that in those years there were no computers in the Observatory (the first one appeared in 1967), 
and the processing was carried out by observers and calculators. 
The processing was carried out, as it was called, ``in two hands'': first by observer, then by calculator. 
Only computation of apparent places of the stars could be made in the Leningrad city computer center. 
Lidia Kostina thoroughly understood the astronomical and computational aspects of this problem and created a new more 
accurate and computationally efficient algorithm of routine computation. 
Using this algorithm, she analyzed the second post-WW2 ZTF-135 series (1955.0--1961.3). 
Later this technique was applied at all Russian latitudinal stations.

Lidia Kostina was also involved in such a complicated task as the compilation of long-term observing programs for 
zenith telescopes, which should be observed for at least 20 years and ensure maximum uniformity of the series by 
observing the same stars. 
She compiled a special program for the new Ulan-Bator latitude station in Mongolia, designed for observations over 
50 years. 
Subsequently, she composed sixth and seventh programs for ZTF-135, which was observed until the end of 2006, that is, 
until the end of the active life of the telescope.

The scientific interest of Lidia Kostina covered many aspects of the Earth's rotation problem. 
Among them, was improvement in the accuracy of computation of the latitudes, and the definition of astronomical 
constants, which she worked with her colleagues. 
In particular, a new definition of the aberration constant was made in 1969, and a new estimation of the principal 
nutation term was obtained in 1988. 
However, she was interested, in the first place, in studying of the long-term changes in the pole motion, such as 
secular, Chandler, and annual terms.

Lidia Kostina, a truly zealous observer, was a living reproach for an observer who was careless or negligent. 
Some observers sometimes called the unstable weather with rapidly changing clouds ``the weather of Lidia Kostina''. 
For its long observing life at ZTF-135 (20.05.1956-28.08.2001), Lidia Kostina obtained 32075 instant latitudes 
during 2279 nights.

Despite being very busy, along with the scientific activities Lidia Kostina has a lot of historical research. 
Their main theme was women of science, women astronomers. 
And no wonder, it was a vivid example of Sophia Vasilievna Romanskaya, the most active observer and 
scientists working in the ZTF-135 group in 1918--1962. 
It was to her that Lidia Kostina's first biographical article was devoted. 
Among about 70 her publications, 12 papers were on the history of astronomy. As a specialist and a person, 
Lidia Kostina enjoyed great respect among colleagues. 
She was very sociable and open, always ready to answer a question, share experiences, interesting information. 
She was an active traveller and often put her notes about an interesting tourist or business trip in the hand-made 
wall newspaper, which was issued at the Observatory in that time.

\begin{figure}
  \begin{minipage}[b]{0.4857\textwidth}
    \includegraphics[clip,width=\textwidth]{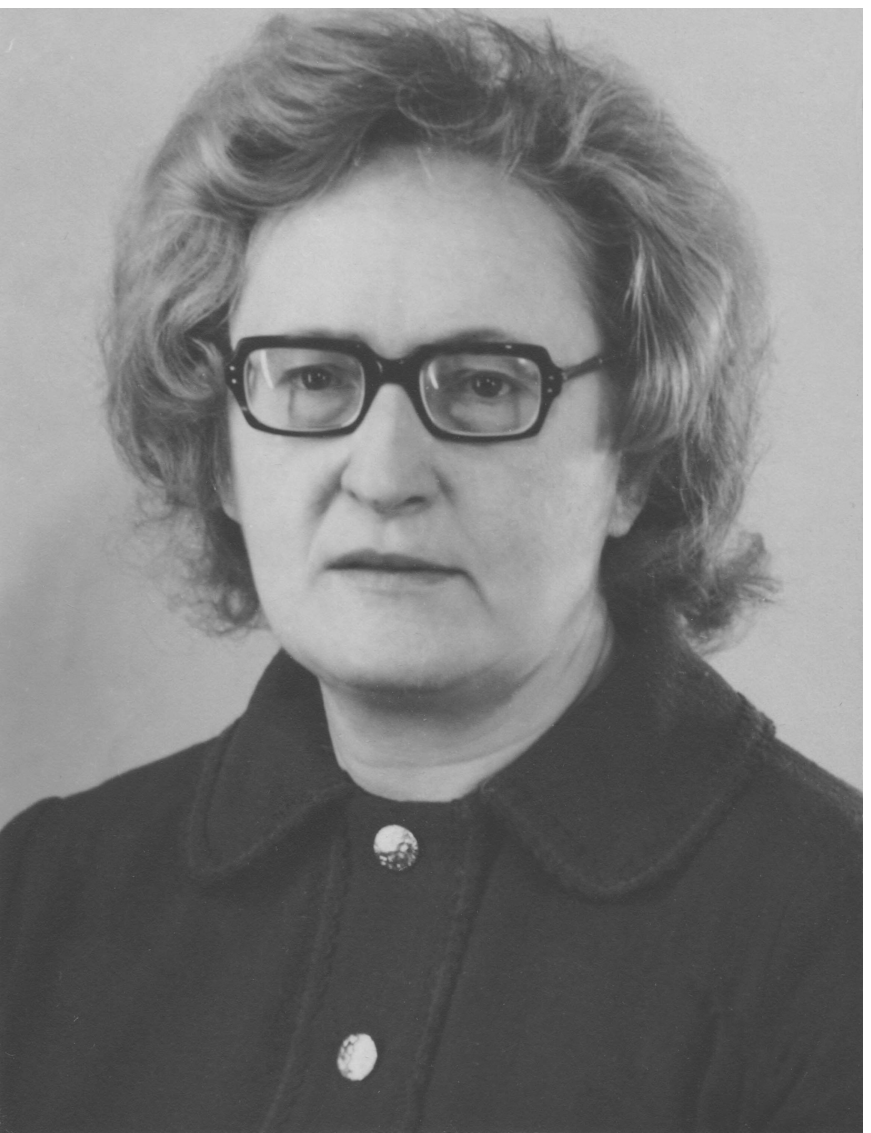}
    \caption{N.R.Persiyaninova, the 1970s.}
  \end{minipage}
  \hfill
  \begin{minipage}[b]{0.4943\textwidth}
    \includegraphics[clip,width=\textwidth]{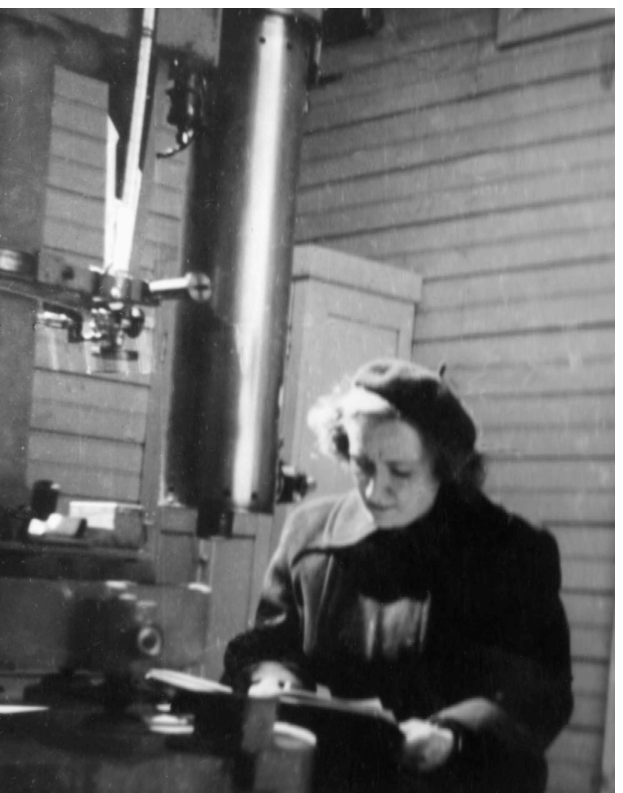}
    \caption{N.R.Persiyaninova at ZTF-135.}
  \end{minipage}
\end{figure}

Natalia Persiyaninova graduated from the astronomical department of the Faculty of mathematics and mechanics 
of the Leningrad State University in 1953 and came to work to the same department of the Pulkovo Observatory. 
Just at that time, preparations for the International Geophysical Year began at the observatory and a new zenith 
telescope ZTL-180 (Zenith Telescope Leningradsky with a lens diameter of 180 mm), manufactured by the Leningrad 
enterprize GOMZ was installed at Pulkovo. 
The first work of Natalia Persiyaninova was devoted to the study of the new instrument. 
She started as a calculator, then, since 1955, she participated in observations at both Pulkovo zenith telescopes, 
but, in the end, she connected her further scientific life with ZTF-135. 

The variety of scientific interests of Natalia Persiyaninova is demonstrated by the list of topics of her publications. 
They include the connection of latitude variations with meteorological factors, non-polar latitude ($z$-term), impact 
of errors in star positions, the determination of the principal term of nutation, investigations of the main components 
in the polar movement, inter-technique comparison of latitude variations. 
In particular, she studied the distribution of clear weather depending on the moon phases.

Everything new that appeared in the mathematical processing of observations, she immediately considered and put into 
practice. 
As soon as the first computer appeared in the observatory in 1967, Natalia Persiyaninova mastered it and provided both 
zenith telescopes with programs for processing of the observations. 
One of the main topics in her work was the application of spectral analysis to the study of variations of latitudes. 
Her dissertation defended in 1969 and devoted on investigation of non-polar variations was one of the first study 
in which correlation and spectral analysis was used for astrometric problems.

Natalia Persiyaninova, as well as Lidia Kostina, always had a high sense of responsibility. 
This was especially true for observations. 
During the period from 08.01.1955 till 16.11.2001 she obtained 33996 instantaneous latitudes during 2414 nights 
(the record for ZTF-135!). 
Numerous observations aimed at the determination of instrument constants, near-pole stars in elongations, scale pairs 
and time series for determining the screw revolution should be added to this value. 
She also took part on latitude observations in Blagoveshchensk (ZTL-180) and Kitab (ZTL-180 and Bamberg ZT).

The scientific authority of Natalia Persiyaninova was very high. The same was her public authority. 
She not only taught students and trainees astrometric wisdom, but also became an adviser in worldly affairs. 
Contacts established during common work continued to exist in correspondence. 
When a scientific conference was held at Pulkovo, discussions in her cozy house were sometimes more active than in and 
around the conference room.

\begin{figure}
  \begin{minipage}[b]{0.512\textwidth}
    \includegraphics[clip,width=\textwidth]{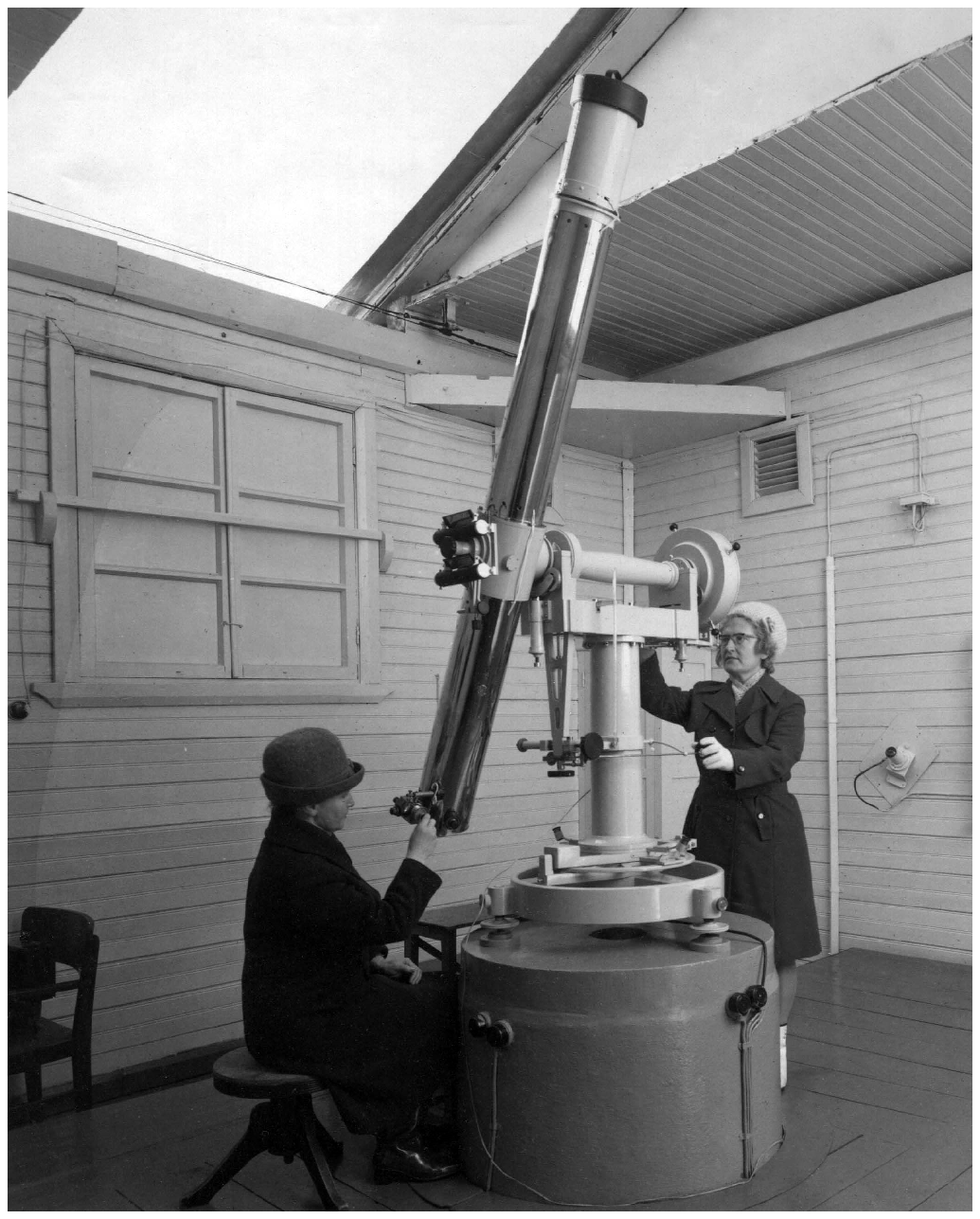}
    \caption{In Pulkovo at ZTF-135.}
  \end{minipage}
  \hfill
  \begin{minipage}[b]{0.468\textwidth}
    \includegraphics[clip,width=\textwidth]{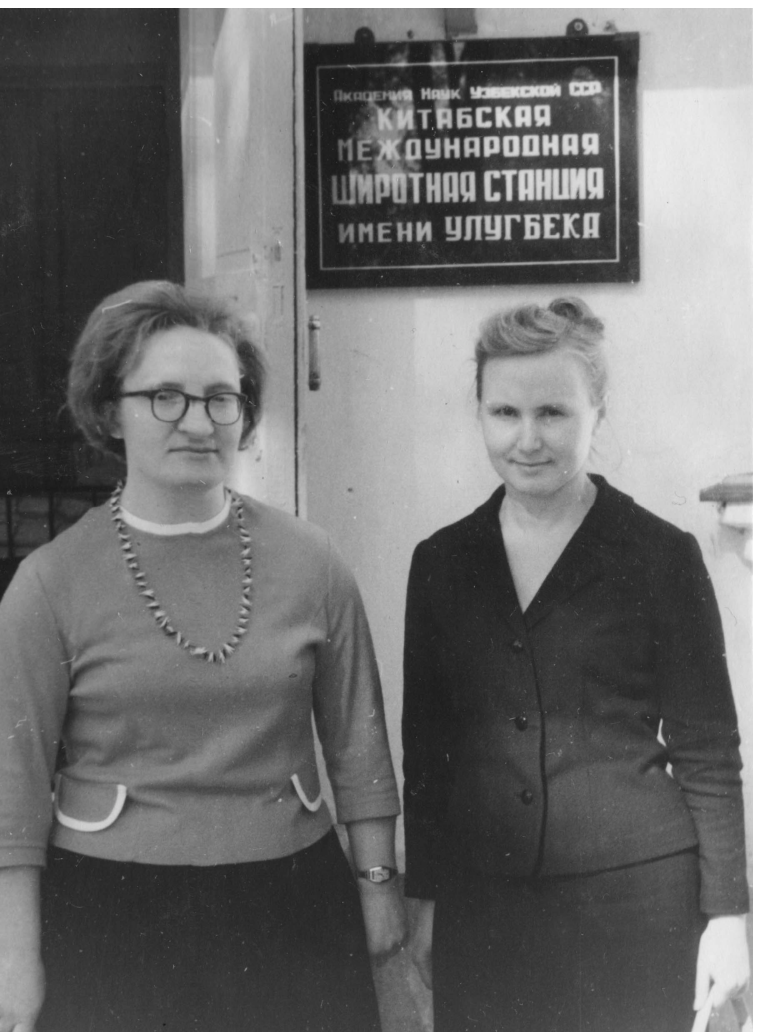}
    \caption{At Kitab ILS stations.}
  \end{minipage}
\end{figure}

Although Lidia Kostina and Natalia Persiyaninova each had their own research topics of interest to them, 
being members of the same Pulkovo latitudinal group, they naturally had many general publications and reports 
at conferences (and in those years astrometric and specialized latitudinal conferences took place in our much more 
often than now), based primarily on observations with ZTF-135. Almost two-thirds of the group's publications were 
written together. 
These are articles on the studies of the of the Pulkovo latitude variations, the instrumental studies, 
the determination of nutation coefficients, the analysis of non-polar variations in latitude observations, 
and the improvement of the catalog of coordinates of latitudinal stars.

Very productive was long-term cooperation of Lidia Kostina and Vladimir Ivanovich Sakharov, many-year head
of the ZTF-135 group. 
They published 17 articles devoted to study of the Chandler, annual and secular components of the pole motion. 
In 1977, they first investigated the relationship between the pole motion and solar activity. 
The presence of common cycles in the variations of the main components of the pole motion and Wolf numbers led them 
to conclude that some of them are generated by solar processes such as 11-yr cycle [1]. 
Studying the Chandler oscillation, the authors were among the first to study the phase changes of this oscillation, 
in particular, around 1925 and drew attention to the fact that neglecting the phase variations can lead to constructing 
of a wrong model of the polar motion. The value of the Chandler period of $1.189 \pm 0.002$~yr obtained by them 
corresponds to the theoretical estimation for the second model of the Earth's structure developed by Michael Molodensky 
(elastic mantle, liquid core with a solid inner core). 
The authors proposed the theory of extreme cycles in the variation of the amplitude of the Chandler wobble. 
Analyzing the maximum amplitudes, they estimated the main low-frequency period in the variations of the Chandler wobble 
of 43.8~yr. Their assumption is that the maximum of the third extreme cycle will fall on 1989, turned out to be close 
to the truth (the maximum was observed in the early 1990s).

Very interesting was also the joint work of Natalia Persiyaninova with Niklen Petrovich Godisov. 
In this study of the main components of the pole motion, the method of sequential separation of components was used 
with parabolic interpolation in the region of the maximum spectral function, which was then used by many authors. 
In the spectrum of amplitudes of the Chandler pole motion, a component with a period of 44~yr was found, 
which corresponds to the conclusions of Sakharov and Kostina.

Over their nearly half-century scientific and, in particular, observing lives, Lidia Kostina and Natalia Persiyaninova 
observed 66071 high-precision latitudes, distributed approximately equally between them. 
A simple estimate (four latitudes per hour in a typical observing program) shows that to obtain such a result, 
each of them spent about 8300 hours with the instrument or about 345 days, i.e. about a whole year!
If one takes into account also auxiliary, but necessary observations, this will be even more than a year.

For preparation of this paper, materials of the archive of the Pulkovo Observatory were used [2--6].


\vspace*{0.7cm}

\noindent{\large\bf REFERENCES}

{

\leftskip=5mm
\parindent=-5mm

\smallskip

1. Kostina L.D., Sakharov V.I. Investigations of the Earth pole motion in the Pulkovo Observatory. In: 150 years of Pulkovo Observatory. Leningrad, Nau-ka, 1989, pp. 137-152.

2. Pulkovo Observatory archive, f. 1, op. 2, d. 105, ll. 112-141 (Personal file of postgraduate student L.D.Kostina).

3. Pulkovo Observatory archive, f. 1, op. 2. d. 863, ll. 34-97 (Personal file of L.D.Kostina).

4. Pulkovo Observatory archive, f, 1. op. 2. d. 385, ll. 163-194 (Personal file of postgraduate student N.R.Persiyaninova).

5. Pulkovo Observatory archive, f. 1, op., 2. d. 905, ll. 178-235 (Personal file of N.R.Persiyaninova).

6. Pulkovo Observatory archive, f. 4, op. 1, d. 106, l. 26 (Photo album ``150 years of Pulkovo Observatory'').

}

\end{document}